\documentclass[
superscriptaddress,
preprint,
amsmath,amssymb,
aip,
jcp,
citeautoscript
]{revtex4-2}

\usepackage{graphicx}
\usepackage{dcolumn}
\usepackage{bm}
\usepackage{braket}
\usepackage[dvipsnames]{xcolor}
\usepackage{soul}
\usepackage{float}

\newcommand{\etal}{{\it~et~al.~}}

\usepackage{subfiles}

\begin{document}

\title{Identifying incoherent mixing effects in the coherent two-dimensional photocurrent excitation spectra of semiconductors}%

\author{Ilaria~Bargigia}
\altaffiliation{Current Address: Center for Nano Science and Technology@PoliMi, Istituto Italiano di Tecnologia, via Pascoli 70/3, 20133 Milano, Italy}
\affiliation{School of Chemistry and Biochemistry, Georgia Institute of Technology, 901 Atlantic Drive, Atlanta, GA~30332, United~States}

\author{Elizabeth~Guti\'errez-Meza}
\affiliation{School of Chemistry and Biochemistry, Georgia Institute of Technology, 901 Atlantic Drive, Atlanta, GA~30332, United~States}

\author{David~A.~Valverde-Ch\'avez}
\affiliation{School of Chemistry and Biochemistry, Georgia Institute of Technology, 901 Atlantic Drive, Atlanta, GA~30332, United~States}

\author{Sarah~R.~Marques}
\affiliation{School of Chemistry and Biochemistry, Georgia Institute of Technology, 901 Atlantic Drive, Atlanta, GA~30332, United~States}

\author{Ajay~Ram~Srimath~Kandada}
\affiliation{Department of Physics and Center for Functional Materials, Wake Forest University, 1834 Wake Forest Road, Winston-Salem, NC~27109, United~States}

\author{Carlos~Silva}
\email{carlos.silva@gatech.edu}
\affiliation{School of Chemistry and Biochemistry, Georgia Institute of Technology, 901 Atlantic Drive, Atlanta, GA~30332, United~States}
\affiliation{School of Physics, Georgia Institute of Technology, 837 State Street, Atlanta, GA~30332, United~States}
\affiliation{School of Materials Science and Engineering, Georgia Institute of Technology, 771 Ferst Drive NW, Atlanta, GA~30332, United~States}

\date{\today}

\begin{abstract}
 We have previously demonstrated that in the context of two-dimensional (2D) coherent electronic spectroscopy measured by phase modulation and phase-sensitive detection, an \emph{incoherent} nonlinear response, due to pairs of photoexcitations produced via linear excitation pathways, contribute to the measured signal as unexpected background [Gr\'egoire et al., J.\ Chem.\ Phys. \textbf{147}, 114201 (2017)]. Here, we simulate the effect of such incoherent population mixing in the photocurrent signal collected from a GaAs solar cell by acting externally on the transimpedance amplifier circuit used for phase-sensitive detection, and we identify an effective strategy to recognize the presence of incoherent population mixing in 2D data. While we find that incoherent mixing is reflected by cross-talk between the linear amplitude at the two time-delay variables in the four-pulse excitation sequence, we do not observe any strict phase correlations between the coherent and incoherent contributions, as expected from modelling of a simple system. 
\end{abstract}

\maketitle


\section{Introduction}

Electronic two-dimensional (2D) spectroscopy is a powerful tool to probe energetic landscapes~\cite{Abramavicius2008,Biswas2022} as it provides relevant information about the coupling between different energy levels, their relaxation dynamics, the coupling between electronic and vibrational levels, and multi-body interactions~\cite{cho2008coherent,Song2014}. While most commonly this technique is implemented through the radiation from a non-linear time-varying coherent polarisation in a phase-matched and time-ordered pulse-sequence 
configuration~\cite{Brida12,Lomsadze2017,Turner2011,Schroter2018}, phase modulation and phase cycling approaches have also been developed~\cite{Tekavec2007,Fuller2015,Karki2022}, with the distinct advantages of allowing for a higher sensitivity and the inspection of longer population times. Furthermore, these approaches are based on the detection of \textit{action} (excitation) signals such as photoluminescence intensity~\cite{Tekavec2007,Lott2011,Perdomo-Ortiz2012,Widom2013,Fuller2015,Gregoire2017,Maly2018,heussman2019measuring,GutierrezMeza2021,Agathangelou2021,Heussman2022,Karki2022}, photocurrent~\cite{Nardin2013,Karki2014,Bakulin2016,Vella2016}, and photoinduced absorption~\cite{Li2016}, and they are therefore suitable for measuring operating devices (e.g.\ solar cells, diodes)~\cite{Nardin2013,Karki2014,Bakulin2016,Vella2016}. The phase-sensitive detection scheme also offers a distinct advantage in the self-stabilization of the experiment. Nevertheless, recent studies with phase modulation and phase cycling approaches have shown that, in condensed matter systems, incoherent signals arising from nonlinear processes can distort and potentially mask the two-dimensional coherent signatures~\cite{Gregoire2017a}. The sought nonlinear coherent signal originates from the interference of the quantum wave-packets generated by the four-pulse sequence. This interference produces a fourth-order excited-state population that ideally undergoes only first-order radiative and non-radiative decay processes. This signal is generally defined as ``coherent" because it directly stems from the four light-matter interactions and it retains a well-defined phase pathway. In condensed-matter systems, however, there is the possibility for the excited-state population to undergo not only first-order decay processes but second-order decay processes as well (i.e.\ bimolecular recombination, exciton-exciton annihilation, auger recombination, photocarrier scattering processes). These processes result in a nonlinear signal which is not coherent in nature, as it is the result of subsequent second-order interactions of the original excited state population produced by each pair of the four-pulse sequence.  It is also important to highlight how emerging/un-optimized devices can show frequency dependent impedance and interface parasitic capacitance that make them nonlinear in nature: nonlinear signal mixing therefore cannot only be an intrinsic characteristic of the material under consideration, but it can also arise from the device architecture as well. As a consequence, extreme care must be taken when interpreting such 2D spectra, but this task is far from trivial. Kalaee et~al.\ proposed a means of distinguishing the coherent from the incoherent contributions~\cite{Kalaee2019} by means of nonperturbative simulation of two-dimensional spectra in a simple two-level molecular system, assuming the presence of exciton-exciton annihilation. These authors showed that when the two excited states have different quantum yields the incoherent contribution acquires a $\pi$ phase shift relative to the coherent signal. In that scenario, it would then be enough to measure the phase of the 2-dimensional coherent signal simultaneously with the phase of the linear signal to allow for the separation of incoherent mutli-body dynamics and coherent excitation pathways.
In our work we build upon this input. We experimentally demonstrate how in a more complex and practical system as a bulk semiconductor the scenario is more convoluted, and separating the coherent from the incoherent component is not straightforward. In fact, contrary to what was predicted by Kalee~et~al. in a very simple model~\cite{Kalaee2019}, we do not observe any strict phase correlations between the coherent and incoherent contributions. However, we show that it is extremely important to acquire and carefully inspect the linear signals in the temporal domain, to identify any presence of incoherent mixing by cross talk of these two signals. As a benchmark, we use a GaAs solar cell and modulate the amount of incoherent mixing by changing the electronic impedance of an external circuit, used to transduce and amplify the photocurrent generated in the solar cell. As we do not affect the physics of the sample itself but we act externally, the homogeneous and inhomogeneous broadening are not perturbed. At the same time, the nonlinear mixing thus produced has the same manifestations as the one generated by incoherent many-body dynamics. Herein we provide an effective tool for recognizing this phenomenon, and discuss a possible way of isolating the coherent response.

\section{Experimental apparatus}

\begin{figure}
    \centering
    \includegraphics[width=8.5cm]{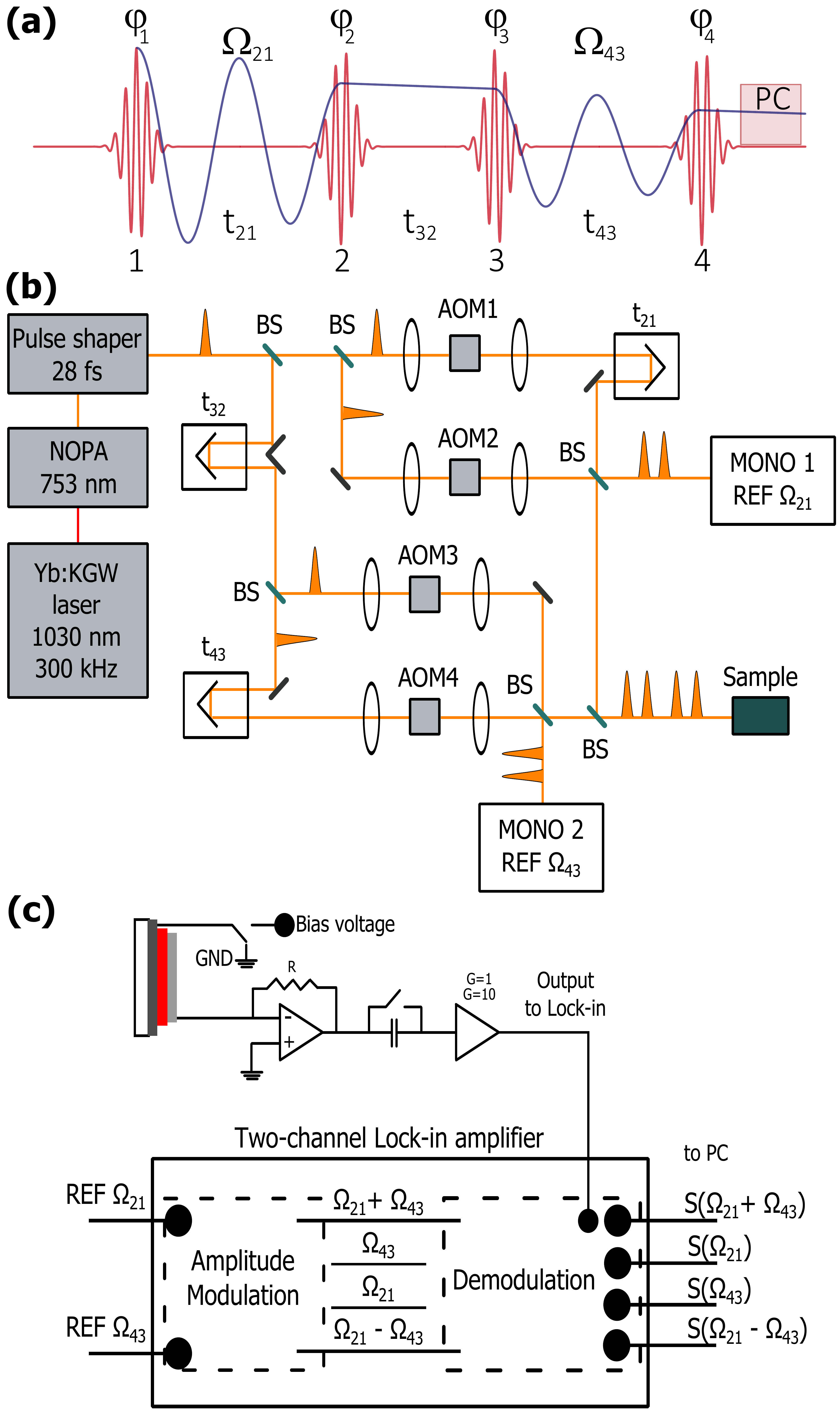}
    \caption{(a) Schematic representation of the pulse sequence used in the experiment. Here, $\phi_i$ represent the phase of each pulse; $t_{ji}$ are the inter-pulse delays, $\Omega_{ji}$ is the frequency at which the phase of each pulse oscillates. PC stands for Photo-Current, the integrated signal that we detect.}(b) Schematic diagram of the set-up for the two-dimensional spectroscopy experiment in collinear geometry. Abbreviations have the following meanings: NOPA, Noncollinear optical parametric amplifier; BS, beam splitter; AOM, acousto-optical modulator (Bragg cell); MONO, monochromator; REF, reference. (c) Schematic of the phase-sensitive apparatus for the photocurrent detection. The top left side represents the sample, connected to the transimpedance amplifier circuitry. The output from the TIA is fed to the digital lock-in amplifier which is able to demodulate the signal at 4 different frequencies. Part (b) is reprinted from ref.~\citenum{GutierrezMeza2021}, Copyright 2021. Authors licensed under a Creative Commons Attribution (CC BY) license.
    \label{fig:setup}
\end{figure}

For a detailed description of the phase modulation scheme adopted in the experiment, please refer to Ref.~\citenum{Tekavec2007}; here we will briefly outline its basic principle of operation. We implement a collinear four-pulse sequence to excite the semiconductor (Fig.~\ref{fig:setup}(a)). The schematic of our setup is reported in Fig.~\ref{fig:setup}(b). The time interval between consecutive laser pulses ($T_{\mathrm{rep}}$) is set by the laser repetition rate, 300$\,$kHz. Out of a single laser pulse-train, the two nested Mach-Zehnder interferometers create four replicas, as reported in Fig.~\ref{fig:setup}(a). By acousto-optic modulation, each of the four replicas undergoes a frequency shift equal to the acoustic frequency $\Omega_i$ of the relative modulator. Although these frequency changes are negligible as compared to the optical frequency, they introduce a shift in the temporal phase of each pulse that oscillates at $\Omega_i$. 
This means that at each laser shot, the experiment is repeated with a sequence of pulses phase-shifted with respect to the previous shot, thus creating two collinear trains of phase-modulated pulse pairs. These two excitation pulse pairs interfere at the sample and produce a population signal oscillating at $\Omega_{21}$ and $\Omega_{43}$, in the kHz range. It can be readily shown that, within this phase cycling scheme, the non-linear signals of interest oscillate at the frequencies $\Omega_{43}-\Omega_{21}$ and $\Omega_{43}+\Omega_{21}$~\cite{Tekavec2007}. In analogy to four-wave mixing experiments, $\Omega_{43}-\Omega_{21}$ and $\Omega_{43}+\Omega_{21}$ are referred to as \textit{rephasing} and \textit{non-rephasing} signals, respectively. These frequency components are extracted simultaneously from the overall \textit{action signal} (be it a photoluminescence, a photo-induced absorption, or a photocurrent signal) by dual lock-in detection. As it is possible to see from Fig.~\ref{fig:setup}(b), optical copies of pulses 1, 2 and 3, 4 are generated at the exit beam splitters of the two twin Mach-Zehnder interferometers, and used to generate the reference signals for the dual lock-in demodulation. The two sets of copies are sent to two monochromators, which spectrally narrow them (thus temporally elongating them), and are then detected by two avalanche photodiodes. The temporal elongation of the pulses provided by the monochromators produces reference signals for time delays $t_{21}$ and $t_{43}$ up to $\sim10$\,ps. It is important to note that, when scanning $t_{21}$, the phase of the references built in this way does not evolve at an optical frequency, but at a reduced frequency given by the difference between the frequency of the signal and that set by the monochromators. This frequency downshift results in an improvement of the signal-to-noise ratio inversely proportional to the frequency downshift itself, which virtually removes the impact of the mechanical fluctuations occurring in the setup on the signals of interest~\cite{Tekavec2007}. In order to obtain the reference signals at the frequencies of the rephasing and non-rephasing signals, $\Omega_{43}-\Omega_{21}$ and $\Omega_{43}+\Omega_{21}$, one of the two photodiode outputs (typically the one at higher frequency) undergoes amplitude modulation (AM) by the output of the other photodiode. The AM signal obtained carries the two sideband frequencies of interest ($\Omega_{43}-\Omega_{21}$ and $\Omega_{43}+\Omega_{21}$) and can then be used for the lock-in demodulation of the \textit{action signal} collected from the sample, as schematically shown in Fig.~\ref{fig:setup}(c) for the specific case of a photocurrent measurement. In our case, the signal is collected from a GaAs solar cell kept under short circuit condition (i.e.\ zero bias) and illuminated by the sequence of the four phase-stabilized laser pulses. The photocurrent signal thus generated is converted into a voltage signal by a transimpedance current amplifier (TIA --- HF2TA Current Amplifier, Zurich Instruments), and then sent to a digital lock-in amplifier (HF2LI Zurich Instruments) for phase-sensitive detection. The 2D maps are built acquiring the demodulated signals at fixed $t_{32}$ times and by scanning $t_{21}$ and $t_{43}$; $t_{21}$ is called the coherence time, $t_{43}$ the detection time, and $t_{32}$ the population waiting time. Specifically, for a given population waiting time, data is sequentially recorded at different coherence times in the interval of interest, typically extending to a few hundred femtoseconds; the detection time is then stepped, and the coherence time scanned repeatedly until the full 2D time response is recorded. Each of such scans produces eight maps, thanks to the capability of the lock-in amplifier to simultaneously demodulate the input signal at multiple frequencies: the in-phase and the in-quadrature maps for the rephasing and non-rephasing frequencies, and the in-phase and in-quadrature ones for the signals oscillating at $\Omega_{43}$ and at $\Omega_{21}$. The maps so obtained in the time domain are, finally, converted to the energy domain by Fourier-transforming the time variables $t_{21}$ and $t_{43}$, and recorded as a function of the population waiting time, $t_{32}$.

\section{Nonlinear incoherent contributions to the 2D lineshape}

\begin{figure*}
    \centering
    \includegraphics[width=17cm]{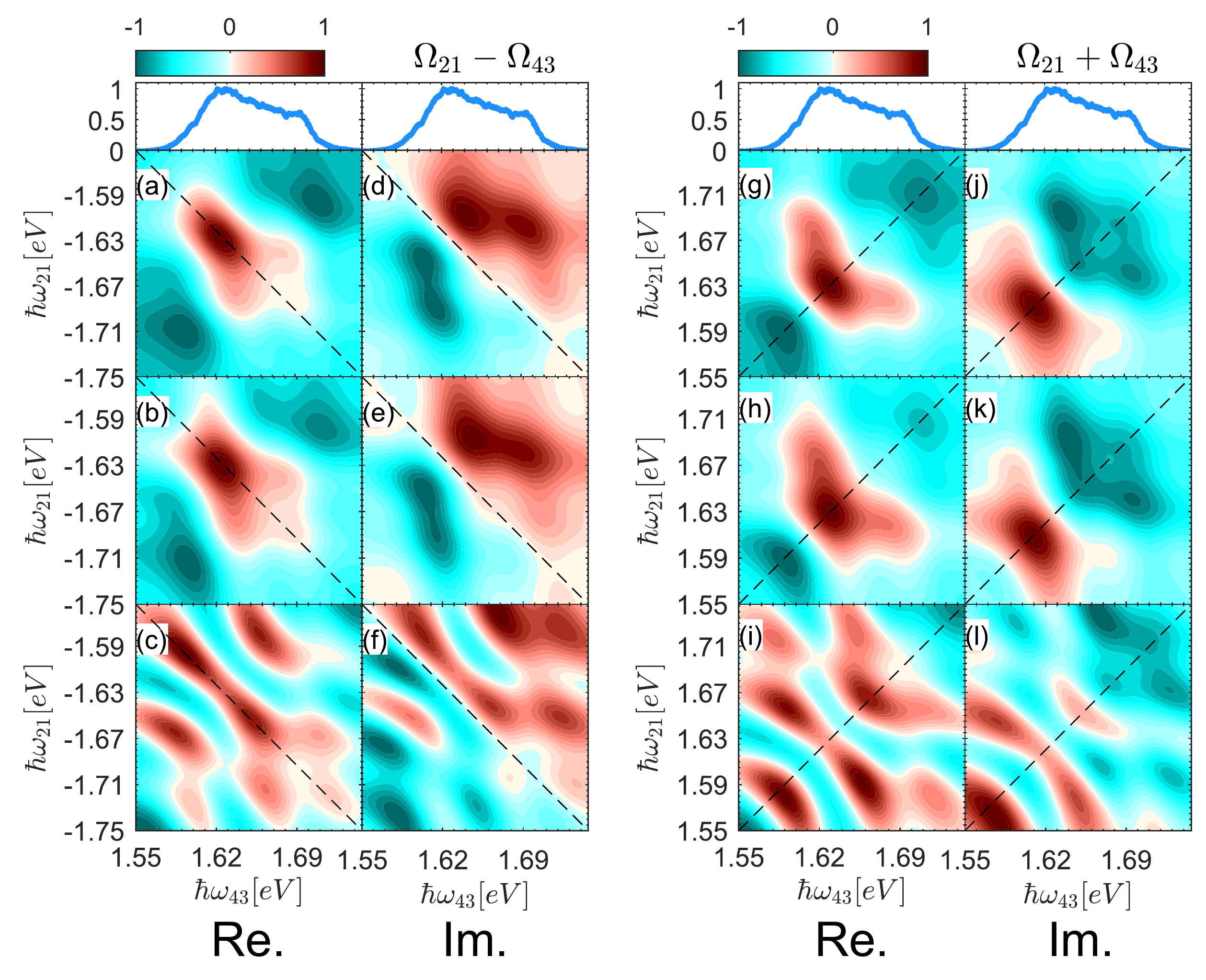}
   \caption{ Two-Dimensional photo-current spectra at a population waiting time of 50\,fs for the three total gains. On the left: real and imaginary components of the rephasing signals for $G_{TOT,1}$ (a,d), $G_{TOT,2}$ (b,e), and $G_{TOT,3}$ (c,f). On the right: real and imaginary parts of the non-rephasing signals for $G_{TOT,1}$ (g,j), $G_{TOT,2}$ (h,k), and $G_{TOT,3}$ (i,l).}
    \label{fig:2D_GaAs_rephasing_nnreph}
\end{figure*}

Our strategy consists of introducing a nonlinearity in the measured signal in a controlled and simple way, by acting on the TIA parameters (an electronic scheme of the circuitry can be found in Fig.~\ref{fig:setup}(c)). The total amplification of the current collected from the device, $G_{TOT}$, depends on the input impedance $R$ of the first operational amplifier and on the voltage gain $G$ of the final operational amplifier: $G_{\mathrm{TOT}} = R \times G$. By increasing the total gain of the TIA we can mimic not only the effect of nonlinear population mixing in the sample but also the nonlinearities arising from external factors such as frequency dependent impedances and parasitic capacitances. 

For this work, we selected three different combinations of $R$ and $G$, as reported in Table~\ref{tab:Gains}. R and G can only assume discrete values. In particular, R can be chosen as: $100$\,V/A, $1$\,kV/A, $10$\,kV/A, $100$\,kV/A, $1$\,MV/A, $10$\,MV/A, or $100$\,MV/A. Instaed, G can either be 1 or 10. The choice of R and G is mainly dictated by the maximum current that the transimpedance amplifier can accept as input, the bandwidth of the signal, and the input impedance. Given that the current coming from the solar cell at the fluence we used is $-3.4 \, \mu$A and that the demodulation frequencies are in the kHz range, the values we chose are all suitable: in fact, above a $G_{TOT}$ of 1MV/A, the TIA is saturated, as its maximum input current becomes equal or lower than $\pm1\,\mu$A. The three combinations of R and G chosen here are those which show the lowest nonlinearity, a medium nonlinearity, and a maximum nonlinearity, while at the same time keeping a good signal to noise ratio and avoiding saturating the TIA. 

\begin{table}
\caption{\label{tab:Gains}Table with the three gains used.}
\begin{ruledtabular}
 \begin{tabular}{p{1cm} p{1cm} c c c c c} 
  &$R$  &$G$ &$G_{TOT}$ &\shortstack{bandwidth\\3dB cut-off} &\shortstack{max\\input\\current\\range} &\shortstack{input\\impedance}\\ 
 \hline 
 $G_{TOT,1}$ & 1\,kV/A  & 1 & 1\,kV/A & 50MHz & $\pm1$\,mA & 50$\Omega$   \\ 
 $G_{TOT,2}$ & 100\,kV/A & 1 & 100\,kV/A  &1.5\,MHz &$\pm10\,\mu$A &100$\Omega$\\
 $G_{TOT,3}$ & 10\,kV/A & 10 & 100\,kV/A &8\,MHz &$\pm10\,\mu$A &50$\Omega$\\ 
\end{tabular}
\end{ruledtabular}
\end{table}

In particular, by changing the total gain, we introduce extra Fourier components in the frequency spectrum of the raw (un-demodulated) output signal of the transimpedance amplifier \textst{, as shown in Fig.}~\ref{fig:LockIn}. (the voltage signal generated at the output of the TIA as it is read by the lock-in amplifier, before any demodulation, is shown in the Supporting Information).   These new frequency components seep in the 2D spectra.    
Fig.~\ref{fig:2D_GaAs_rephasing_nnreph} shows the real (left column) and imaginary (right column) components of the rephasing and the nonrephasing maps for a population waiting time of $t_{32} = 50$\,fs. Maps (a), (d), (g) and (j) were collected with $G_{TOT,1}$; maps (b), (e), (h) and (k) were collected with $G_{TOT,2}$; maps (c), (f), (i) and (l) were collected with $G_{TOT,3}$.
Rephasing and nonrephasing maps taken with the lowest total gain $G_{\mathrm{TOT,1}}$, (a),~(d)/(g),~(j), show a primarily absorptive lineshape: both real and imaginary parts of the signal are dominated by a strong resonant feature lying on the diagonal and roughly centered on the laser central energy $\omega_L = 1.65$\,eV. When the input impedance of the TIA circuit is increased and the total gain is $G_{TOT,2}$, slight changes appear in the rephasing and nonrephasing maps as the spectral features lose resolution. The nonrephasing signal seems to be more strongly affected: while the main positive feature in the real part of the signal remains substantially unaffected, the relative weight of the negative peaks is now more strongly favouring the lower energy feature. The maximum nonlinearity is seen for maps (c),~(f)/(i),~(l), recorded for $G_{TOT,3}$. Here the nonlinear signal induces a plethora of new features, both positive and negative and both on and off the diagonal. While, if taken singularly, it is difficult to recognize the presence of incoherent mixing in these set of maps, the time-domain maps of the \textbf{linear signals} oscillating at $\Omega_{21}$ and at $\Omega_{43}$ provide a clear indication of this phenomenon, as detailed in the following section. 

\subsection{Analysis of the linear time-domain measurement}
\begin{figure}
    \centering
    \includegraphics[width=8.5cm]{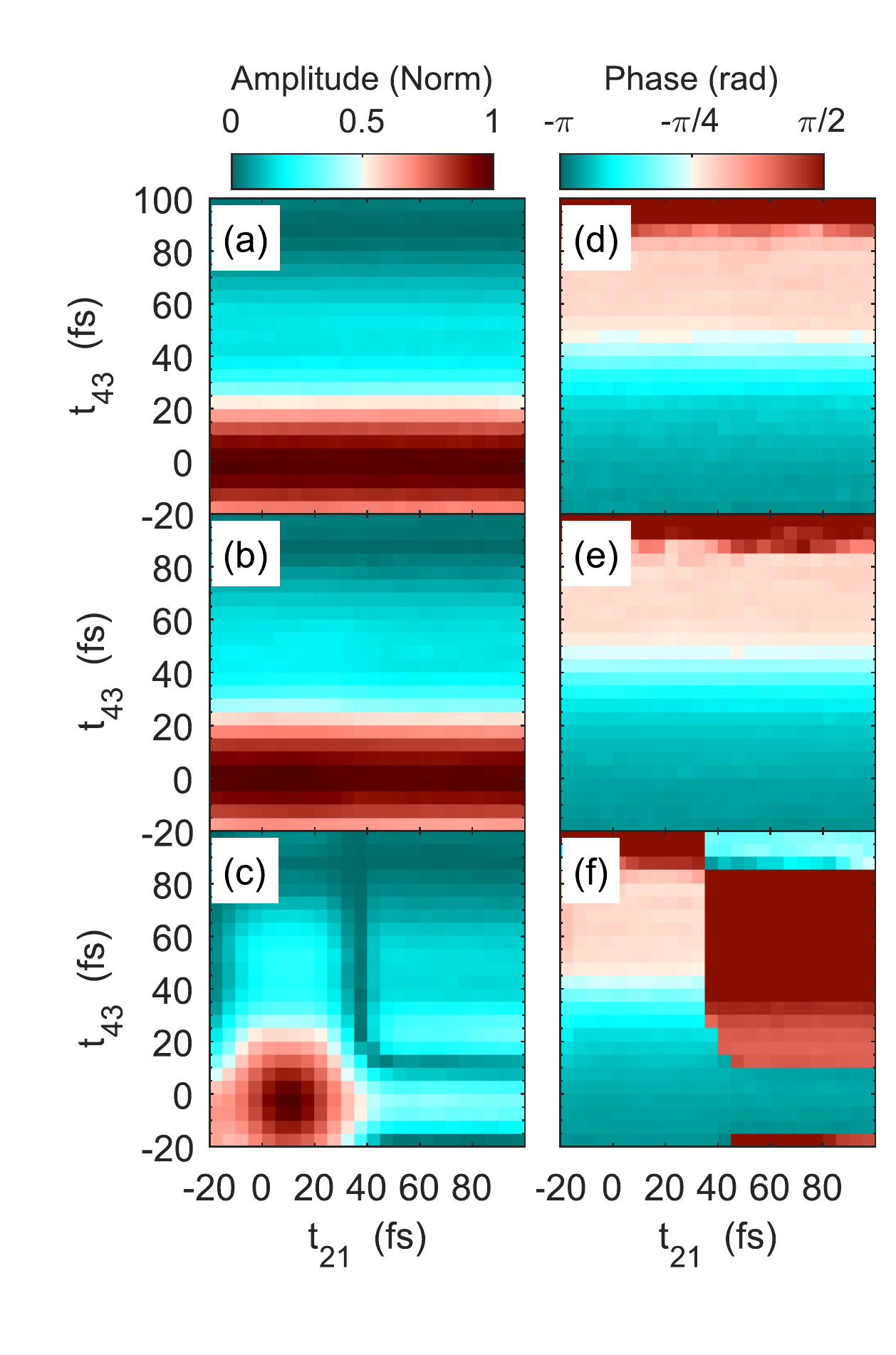}
    \caption{Two-dimensional photo-current data in the time-domain, demodulated at $\Omega_{43}$. }(a)--(c): Amplitude of the linear signal demodulated at $\Omega_{43}$ for $G_{TOT,1}$, $G_{TOT,2}$, and $G_{TOT,3}$ respectively. (d)--(f): Phase of the linear signal demodulated at $\Omega_{43}$ for $G_{TOT,1}$, $G_{TOT,2}$, and $G_{TOT,3}$.
    \label{fig:linear_exp_time}
\end{figure}

\begin{figure*}
    \centering
    \includegraphics[width=14cm]{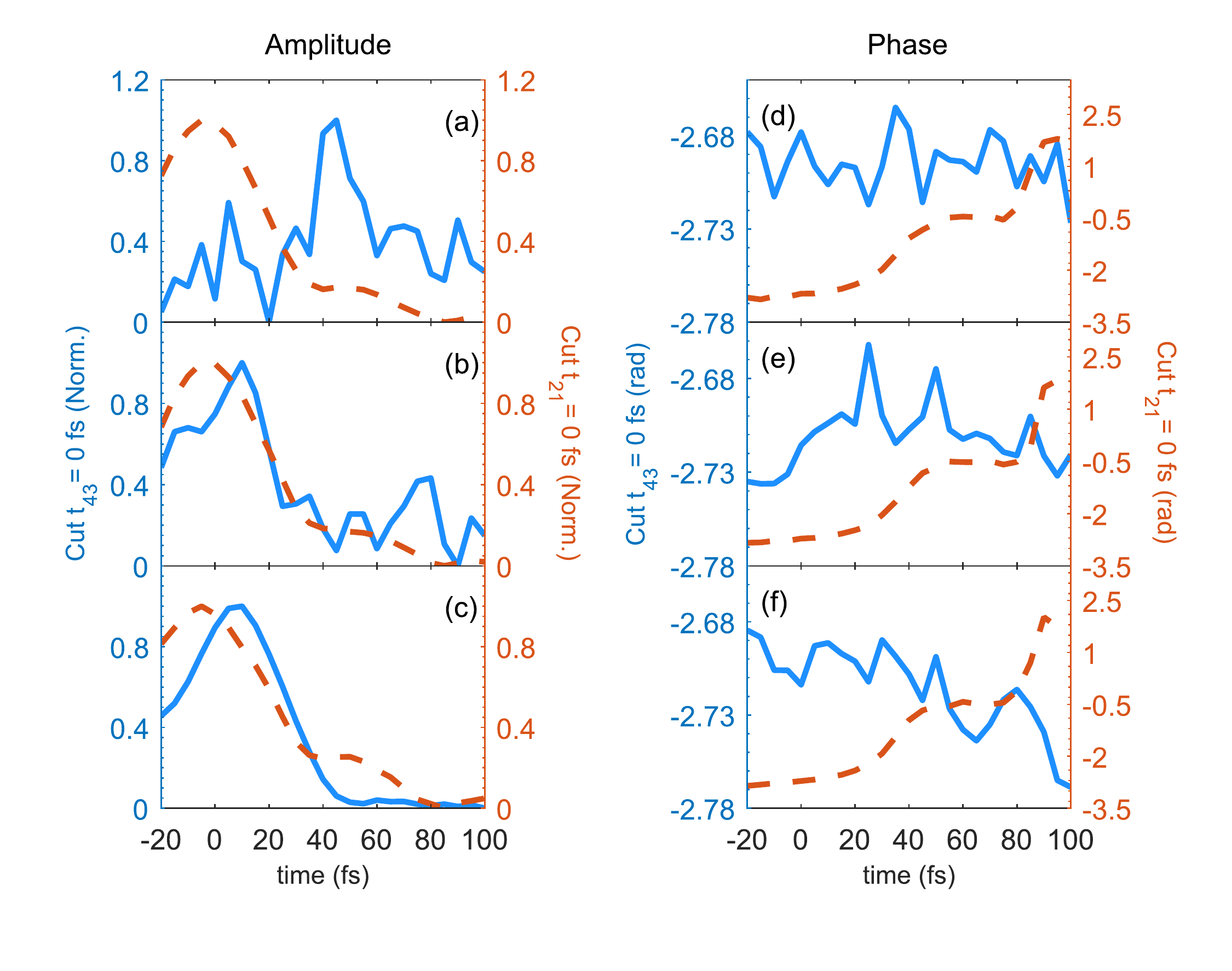}
    \caption{(a)--(c): Cuts of the amplitude maps along the $t_{43}$ temporal scan direction ($t_{43}=0$\,fs, blue, left axis) and the $t_{21}$ direction ($t_{21}=0$\,fs, red, right axis) for $G_{TOT,1}$, $G_{TOT,2}$, and $G_{TOT,3}$ respectively, for the signal demodulated at $\Omega_{43}$. (d)--(f): Cuts of the phase maps along the $t_{43}$ temporal scan direction ($t_{43}=0$\,fs, blue, left axis) and the $t_{21}$ direction ($t_{21}=0$\,fs, red, right axis) for $G_{TOT,1}$, $G_{TOT,2}$, and $G_{TOT,3}$ respectively, for the signal demodulated at $\Omega_{43}$.}
    \label{fig:linear_exp_time_cuts}
\end{figure*}

Careful inspection of the time-domain linear signals demodulated at $\Omega_{43}$ and $\Omega_{21}$ shows a dependence of the signal distribution on the total gain of the TIA, as exemplified in Fig.~\ref{fig:linear_exp_time} for the normalized amplitude (left column) and phase (right column) of the signal demodulated at $\Omega_{43}$. Let us first examine the data for the lowest gain, Fig.~\ref{fig:linear_exp_time} (a),(d). Because we are demodulating at $\Omega_{43}$, the amplitude is expected to oscillate along the $t_{43}$ axis (for a fixed value of $t_{21}$) and remain constant along the $t_{21}$ axis (for a fixed value of $t_{43}$). On the other hand, because our reference frequency is far from the transition resonance we are scanning, the phase shows a staircase-like behaviour along $t_{43}$, whose periodicity is related to the total bandwidth of the response. This staircase behaviour has been already observed in a model quantum system, atomic $^{87}$Rb vapor, when the laser pulse was tuned such that the two excited state amplitudes of the D transition, $|a_1|$ and $|a_2|$, were excited with increasingly equal weight~\cite{Tekavec2006}. As expected, the phase assumes a constant value along $t_{21}$. By changing $G_{TOT}$, we introduce a non-linear component: the maps acquired for $G_{TOT,3}$ show the highest degree of nonlinearity both in the amplitude and in the phase. This behaviour is more evident if we take cuts along the two orthogonal axis $t_{21}=0$\,fs and $t_{43}=0$\,fs of both amplitude and phase, Fig.~\ref{fig:linear_exp_time_cuts}. To better compare the amplitude cuts along the two orthogonal directions and highlight the shape of the curves, data have been normalized at 1. For $G_{TOT,1}$ we indeed see that the amplitude of the signal oscillates only along the $t_{21}=0$\,fs direction (dashed orange line), while it remains constant for the cut along $t_{43}=0$\,fs (solid blue line). As $G_{TOT}$ is varied, the amplitude of the signal starts oscillating along $t_{43}=0$\,fs as well, with a certain time delay as it is especially evident for  $G_{TOT,3}$ (panel (c)). This is indicative of cross-talk between the two linear channels due to nonlinear signal contributions in the amplification circuit. In this extreme case, also the phase seems to be affected, albeit in a lower measure. Comparing the last two rows, which represent data acquired at the same $G_{TOT}$, G seems to have a greater influence on the amount of nonlinearity introduced in the system compared with R.  

These data can be interpreted in the light of the formalism developed by Tekavec~\etal where a two-pulse sequence originating from a single Mach-Zehnder interferometer was interacting in the sample~\cite{Tekavec2006}. For each value of the inter-pulse delay $t_{ji}$, the photocurrent signal gets demodulated by the lock-in amplifier at one of the phase modulation reference frequencies $\Omega_{ji}$ ($j = 2$ or 4 and $i = 1$ or 3 depending on which of the two internal interferometers is being measured). Then, the lock-in amplifier multiplies the alternating-current (ac) response at a particular inter-pulse delay $\mathcal{S}(t_{ji},t^{\prime})$ by a reference waveform $\mathcal{R}(t_{ji},t^{\prime},\theta_{\mathrm{ref}})$, and then removes the ac components via a low-pass filter with characteristic time constant $\tau_{\mathrm{int}}$, which limits the integration time of the measurement. Here the time variable $t^{\prime}$ is a discrete temporal variable accounting for the period of the many pulse-sequence repetitions over the integration window limited by $\tau_{\mathrm{int}}$. The constant reference phase that includes the spectral phase difference between the two pulses is $\theta_{\mathrm{ref}}$. We measure the in-phase ($X$, nominally with $\theta_{\mathrm{ref}\,X} = 0^{\circ}$) and in-quadrature ($Y$, $\theta_{\mathrm{ref}\,Y} = 90^{\circ}$) components of the time-integrated photocurrent~\cite{Tekavec2006}:
\begin{align}
    \begin{split}
     X(t_{ji}) &= \frac{1}{\tau_{\mathrm{int}}} \int_0^{\infty} dt^{\prime}\,\mathcal{S}(t_{ji},t^{\prime}) \mathcal{R}(t_{ji},t^{\prime},\theta_{\mathrm{ref}\,X})e^{-t^{\prime}/\tau_{\mathrm{int}}} \\
     Y(t_{ji}) &= \frac{1}{\tau_{\mathrm{int}}} \int_0^{\infty} dt^{\prime}\,\mathcal{S}(t_{ji},t^{\prime}) \mathcal{R}(t_{ji},t^{\prime},\theta_{\mathrm{ref}\,Y})e^{-t^{\prime}/\tau_{\mathrm{int}}},
     \end{split} 
     \label{eq:X&Y}
\end{align}
such that the total vectorial output of the amplifier is $Z(t_{ji}) = X(t_{ji}) + iY(t_{ji})$. The reference waveform is the interference contribution to the pulse-pair power spectrum~\cite{Tekavec2006}:
\begin{equation}
    \mathcal{R}(t_{ji},t^{\prime}) \propto \cos \left[ \omega_{\mathrm{ref}}t_{ji} - \Omega_{ji}t^{\prime} - \theta_{\mathrm{ref}} + \theta_{\mathrm{0}} \right],
    \label{eq:ref}
\end{equation}
where $\omega_{\mathrm{ref}}$ is the central frequency determined by the monochromator MONO 1(2) in Fig.~\ref{fig:setup}(a), and $\theta_0$ is an arbitrary phase offset imposed by the lock-in amplifier. In the absence of an incoherent nonlinearity in the material response, and solely contribution from the linear population $n^{(2)}$ that is due to the two-pulse sequence, the purely coherent response in a 3D semiconductor $\mathcal{S}_{\mathrm{coh}}$ is given by~\cite{Gregoire2017a}

\begin{widetext}
    \begin{align}
        \begin{split}
        \mathcal{S}_{\mathrm{coh}}(t_{ji}) &\propto f(t_{ji}) \otimes n^{(2)}(t_{ji});\\
        n^{(2)} &= \frac{2}{\hbar^2} \Re \int_\mathbb{R} \mid \mu_{eg} \mid^22\pi \left( \frac{2m^*}{\hbar^2} \right)^{3/2}  \sqrt{\hbar\omega-E_g} \quad \alpha_i(\omega) \alpha_j(\omega) e^{-i \omega t_{ji} - \Gamma_{eg}t_{ji}} e^{i\phi_{ji}} d\omega,
        \end{split}
    \label{eq:S_coh}
    \end{align}

where $f(t_{ji})$ is the instrument response function determined by the temporal pulsewidth, which must be convoluted with $n^{(2)}$. In equation~\ref{eq:S_coh}, $\mu_{eg}$ is the ground-to-excited-state transition dipole moment, and is multiplied by the joint density of states for a 3D semiconductor with energy gap $E_g$; $\alpha_{i(j)}$ is the spectral amplitude of the \textit{i}th (\textit{j}th) pulse, $\Gamma_{eg}$ is the dephasing rate of the transition, and $\phi_{ji}$ is the instantaneous phase difference between the two excitation pulses, which is modulated at frequency $\Omega_{ji}$. For spectrally identical, transform limited pulses and assuming the reference phase from the lock-in to be zero, the instantaneous phase difference is simply $\phi_{ji} =  \Omega_{ji}  t' + (\theta_{2}(\omega)-\theta_{1}(\omega)) + \theta_{0}= \Omega_{ji}  t'$, $\Omega_{ji}$ being the difference in the frequencies of the two acousto-optic modulators.
Inserting equation~\ref{eq:S_coh} into equation~\ref{eq:X&Y}, we can find the expressions for $X(t_{ji})$ and $Y(t_{ji})$:

\begin{align}
    \begin{split}
     X(t_{ji}) = f(t_{ji}) \otimes & \int_\mathbb{R} d\omega \mid \mu_{eg} \mid^2 2\pi \left( \frac{2m^*}{\hbar^2} \right)^{3/2}  \sqrt{\hbar\omega-E_g} \quad
     \alpha_i(\omega) \alpha_j(\omega) \\ &\times \cos \left[ - (\omega - \omega_{\mathrm{ref}})t_{ji}  \right]
     e^{(-\Gamma_{ji}t_{ji})};\\
     Y(t_{ji}) = f(t_{ji}) \otimes & \int_\mathbb{R} d\omega \mid \mu_{eg} \mid^2 2\pi \left( \frac{2m^*}{\hbar^2} \right)^{3/2}  \sqrt{\hbar\omega-E_g} \quad\alpha_i(\omega) \alpha_j(\omega) \\ &\times \sin \left[ - (\omega - \omega_{\mathrm{ref}})t_{ji}  \right] e^{(-\Gamma_{ji}t_{ji})}. 
     \end{split} 
     \label{eq:XYDemod}
\end{align}
\end{widetext}

\begin{figure}
    \centering
    \includegraphics[width=6 cm]{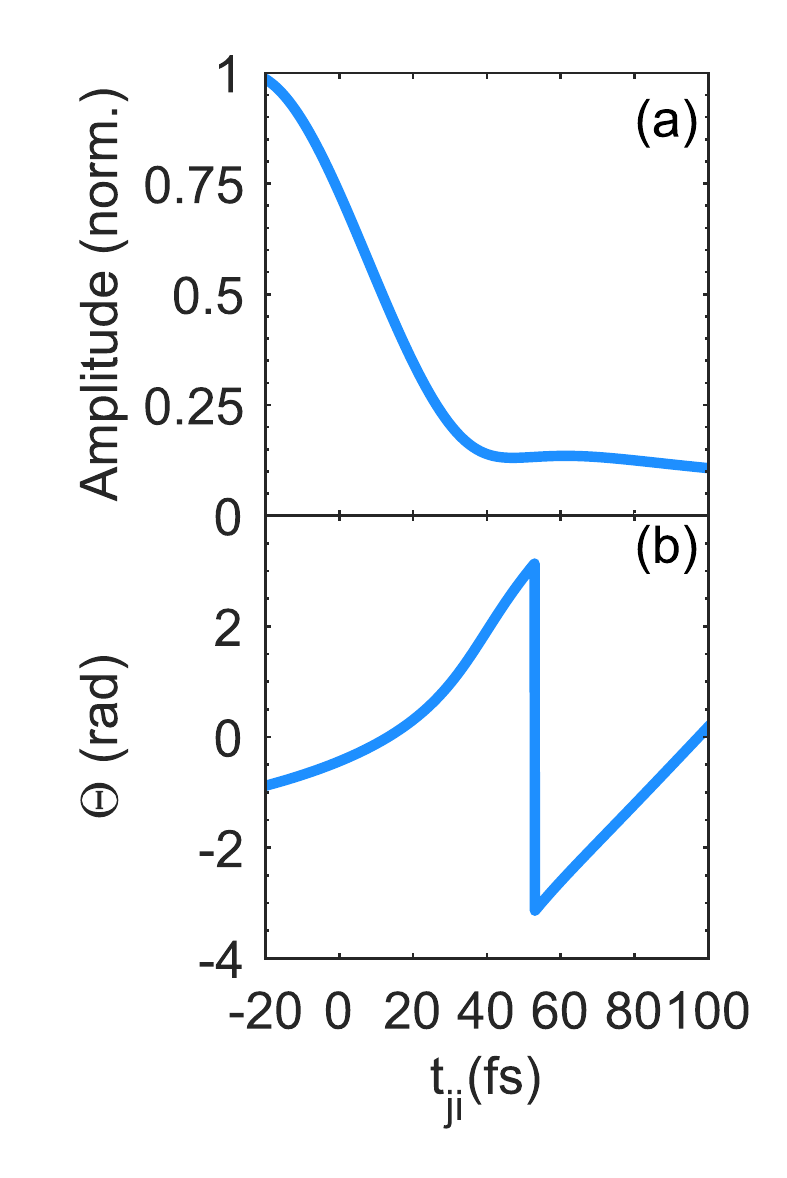}
    \caption{(a): Simulated amplitude of the total signal $Z(t_{ji}) = X(t_{ji}) + iY(t_{ji})$, obtained using equation \ref{eq:XYDemod} for a two-pulse experiment}. (b): Simulated phase of $Z(t_{ji}) = X(t_{ji}) + iY(t_{ji})$, obtained using equation \ref{eq:XYDemod} for a two-pulse experiment.
    \label{fig:Simulation_Lin}
\end{figure}

Using equation \ref{eq:XYDemod} we can simulate the amplitude $A(t_{ji}) =\sqrt{X^2(t_{ij}) + Y^2(t_{ij})}$ and the phase $\Theta(t_{ij})$ of the signal acquired in the specific case of GaAs (see SI for details), and compare them with the experimental data: results are reported in Fig.~\ref{fig:Simulation_Lin}(a) and (b). We can see that there is good qualitative agreement with the data characterized by the least contribution of nonlinear mixing: the amplitude is a slowly decaying function, while the phase is a monotonically non-decreasing function. There are, however, aspects of the simulations which pertain to the modeling of the semiconductor that we are not able to properly capture, for example our simulations show a sharp $2\pi$ jump in the phase. As it is not the scope of this paper the modeling of a layered GaAs solar cell, we will not study this behaviour in detail. 
We explicitly note that this formalism does not show any dependence of the signal demodulated at $\Omega_{43}$ on $t_{21}$: this stems from the fact that it was specifically developed without taking into account the effects of nonlinear mixing which we are discussing here and, as we are showing in Fig.~\ref{fig:linear_exp_time}, do cause a cross-talk between the two temporal axis.

\subsection{Analysis of the time-domain four-pulse measurement}
\begin{figure}
    \centering
    \includegraphics[width=8.5cm]{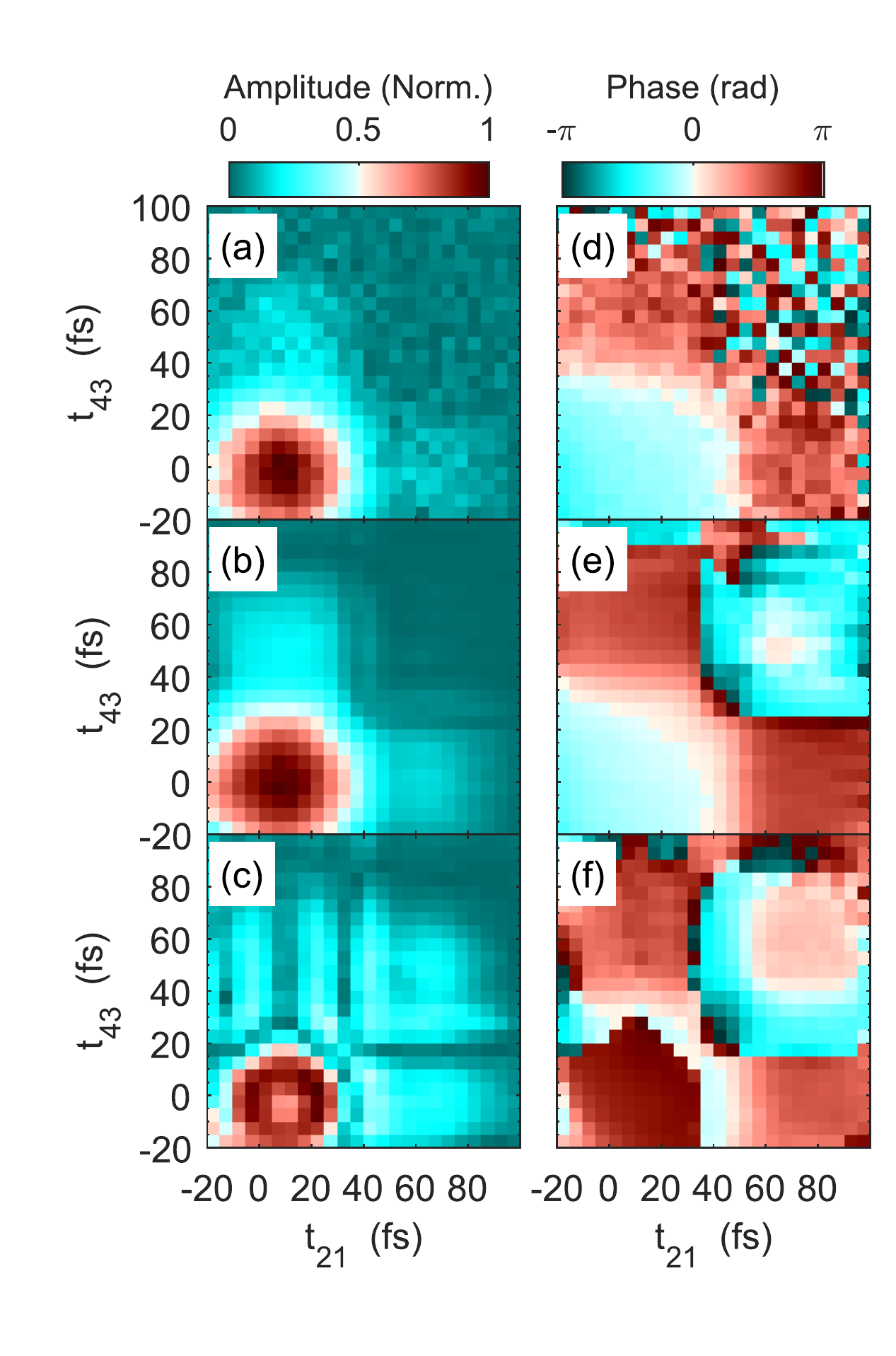}
    \caption{(a)--(c): Amplitude of the nonrephasing signal demodulated at $\Omega_{43} + \Omega_{21}$ for $G_{TOT,1}$, $G_{TOT,2}$, and $G_{TOT,3}$ respectively. (d)--(f): Phase of the nonrephasing signal demodulated at $\Omega_{43} + \Omega_{21}$ for $G_{TOT,1}$, $G_{TOT,2}$, and $G_{TOT,3}$.}
    \label{fig:nonreph_data_time}
\end{figure}

The time-domain phase and amplitude maps corresponding to the 2D spectra for the nonrephasing signal are reported in Fig.~\ref{fig:nonreph_data_time} (see SI for the rephasing maps). In both cases, the gain has a strong impact on the amplitude spectrum, whereas the phase is mostly unaffected by the nonlinearity introduced in the system. The different noise level observed between the different maps is to be expected, as the gain of the conversion from photocurrent to voltage signal is progressively changed. While the time-domain maps of the linear signals oscillating at $\Omega_{21}$ and at $\Omega_{43}$ clearly show signs of nonlinear mixing, it is more difficult to discern any hint of this phenomenon in the time-domain data for the nonrephasing and rephasing signals. We can write the total coherent response $\mathcal{S}_{\mathrm{coh}}$ as:

\begin{equation}
    \begin{split}
        \mathcal{S}_{\mathrm{coh}} \propto f(t_{ji}) \otimes \big[ n^{(2)}(t_{ji}) &+ n^{(2)}(t_{lk}) \\
        &+ n^{(4)}(t_{ji},t_{lk};t_{kj}) \big],
    \end{split}
    \label{eq:S_tot}
\end{equation}

where $n^{(4)}(t_{ji},t_{lk};t_{kj})$ is the fourth order population responsible for the coherent response of the material:

\begin{equation}
    \begin{split}
    n^{(4)}(t_{ij},t_{kl};t_{kj}) &= S_4(t_{43},t_{21}) +S_1(t_{43},t_{21},t_{32})  \\ 
        &+ S_3(t_{43},t_{21})+S_2(t_{43},t_{21},t_{32}) \\
    &= \braket{\psi_4|\psi_{321}} + \braket{\psi_{432}|\psi_1} \\
    &+ \braket{\psi_{421}|\psi_3} + \braket{\psi_{431}|\psi_2}.
    \end{split}
\end{equation}

$S_4(t_{43},t_{21})$ and $S_1(t_{43},t_{21},t_{32})$ are the nonrephasing components, whereas $S_3(t_{43},t_{21})$ and $S_2(t_{43},t_{21},t_{32})$ are the rephasing parts of the signal.
Our detected signal is the result of the superposition of the coherent response of the GaAs solar cell $\mathcal{S}_{\mathrm{coh}}$ and a nonlinear contribution. In a 3D semiconductor as GaAs, it is not that trivial to isolate these two components.  

\section{Discussion}
\begin{figure}
    \centering
    \includegraphics[width=8.5cm]{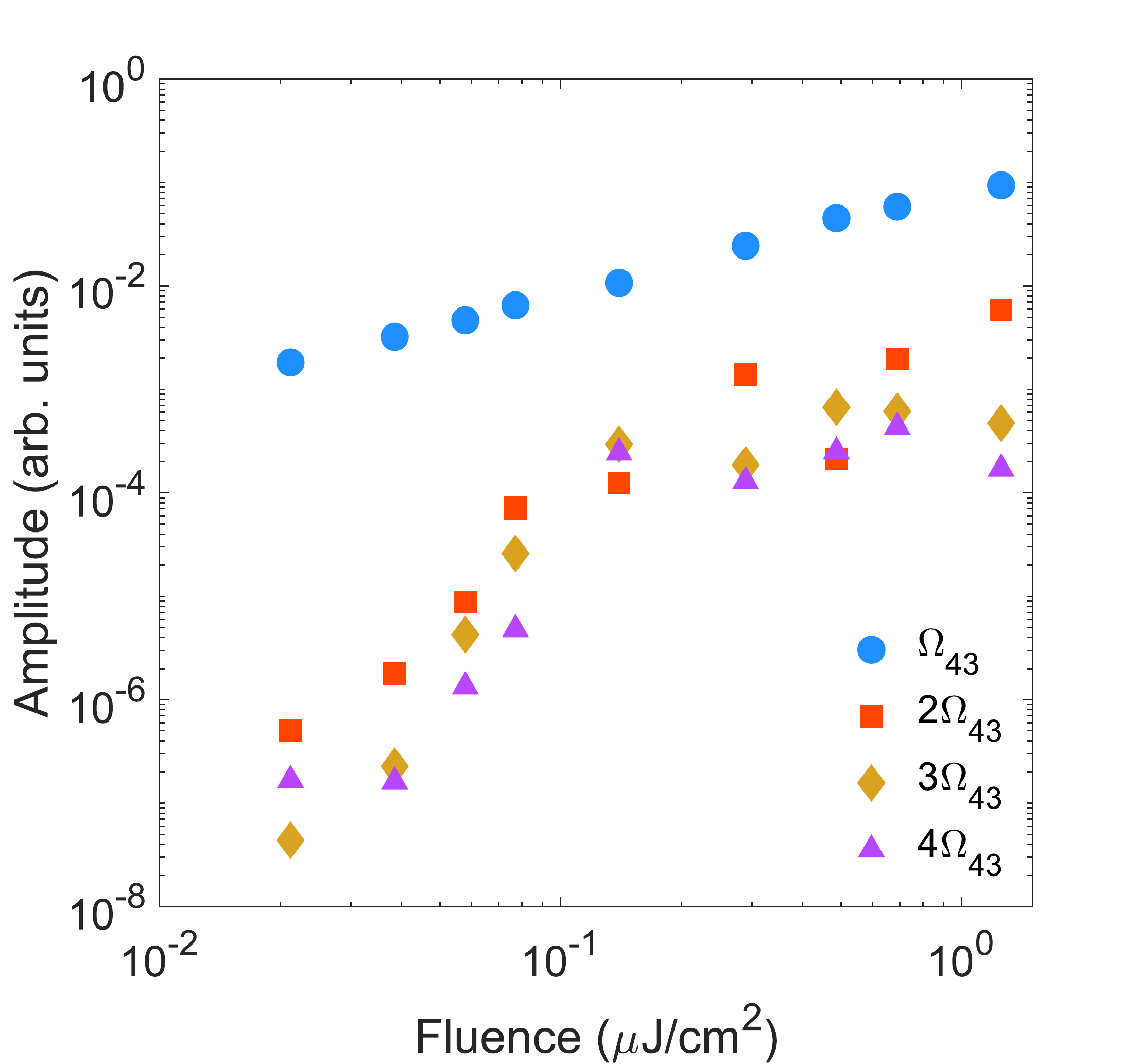}
    \caption{Fluence dependence of the photocurrent signal collected in a two-pulse experiment and demodulated at frequencies \textit{m}$\Omega_{43}$, \textit{m}=1-4.}
    \label{fig:NHarmonics}
\end{figure}

Incoherent population mixing can heavily impair the correct interpretation of 2D spectra of condensed-matter systems. It is therefore important to recognize when this effects are predominant and when we can instead neglect them. We recently discussed the presence of Frenkel biexcitons in a model of a hybrid HJ photophysical aggregate polymer using 2D photoluminescence excitation spectroscopy.\cite{GutierrezMeza2021} In that context, the analysis of the amplitude temporal spectra showed no cross-talk between the two time axis, thus ruling out incoherent mixing effects. The reason behind this is that the fluence was low enough not to generate substantial interparticle interactions. As a general rule of thumb, in fact, nonlinear population dynamics become sufficiently important to mask the true coherent response of the material the more photoexcitations are mobile and subjected to reciprocal interactions (e.g. scattering processes, Auger recombination, bimolecular annihilation processes). While the effort of discerning the coherent signal from the incoherent one based on the assumption that they show a definite phase difference has shown some potential in theoretical simulations of a specific two-level molecular system,\cite{Kalaee2019} this approach may not be applicable in general. In this work we introduce different degrees of nonlinearity by changing the electronic impedance of the external transduction circuitry. By using the formalism developed by Tekavec et al.\cite{Tekavec2006} we are able to appropriately simulate the coherent photocurrent signal originating from a two-pulse experiment: discrepancies between our simulations and the data can be attributed to our assumption of a parabolic density of states, where in actuality the system has a non-parabolic dispersion. When we introduce higher nonlinearities in the system, however, there is no apparent simple link between the phases of the coherent and incoherent contributions. The results obtained by Kalaee et al.\cite{Kalaee2019} are based on the assumption of a simplistic three-level system where the only means of recombination is exciton-exciton annihilation. In real-world, complicated systems the complexity of the interactions require a different approach to the problem. Inspired by the work developed by Grégoire et al.\cite{Gregoire2017a}, where they assumed a weak bimolecular annihilation and the time-integrated signal had a quadratic dependence from the initial population density, we expand $\mathcal{S}_{\mathrm{tot}}$ as a power series, where $\beta_i$ are constants:
\begin{equation}
    \begin{split}
        \mathcal{S}_{\mathrm{tot}} \sim \beta_0 + \beta_1 S_{coh}^1 + \beta_2 S_{coh}^2 + \mathcal{O}(S_{coh}).
    \end{split}
    \label{eq:ExpansionSeries}
\end{equation}
As a first approximation, we can limit the expansion at the second order. $S_{coh}^2$ is the time-varying function that accounts for the incoherent population contributions. 
We can expand $S_{coh}^2$ by squaring the terms that appear in Equation~\ref{eq:S_tot}, thus obtaining a series of terms that oscillate at various frequencies (see SI for details). Some of these oscillation frequencies correspond to the sum frequency $\Phi_{43}+\Phi_{21}$ and to $\Phi_{43}$:
\begin{widetext}
    \begin{gather}
        2 S_{21} (t_{21},\Phi_{21}) \cdot S_{43} (t_{43},\Phi_{43}) = 8\Re \sum[|\mu_{ag}|^2 \alpha^2 (\omega_{ag})]^2  
         e^{-i \omega_{ag} (t_{43}+t_{21})} e^{i \Phi_{SUM}}\\
        2 S_{21} (t_{21},\Phi_{21}) \cdot S_{3} (t_{43}, t_{21},\Phi_{DIFF}) = 8 \Re \sum |\mu_{ag}|^4 \mu_{bg}^2
        \alpha^4(\omega_{ag}) \alpha^2 (\omega_{bg}) e^{-i\omega_{ag}(t_{43}+t_{21})} e^{i\omega_{bg} t_{21}} e^{i\Phi_{43}}\\
        2 S_{21} (t_{21},\Phi_{21} )\cdot S_2 (t_{43},t_{21},\Phi_{DIFF}) =  
        8 \Re \sum |\mu_{ag}|^4 
        \mu_{bg}^2 \alpha^4 (\omega_{ag}) \alpha^2 (\omega_{bg})\\ \times e^{-i(\omega_{ag}-\omega_{bg})t_{21}} 
         e^{-i\omega_{ag}t_{43} -i (\omega_{ag} -\omega_{bg}) t_{32}} e^{i\Phi_{43}}. \nonumber
    \end{gather}
\end{widetext}
These terms oscillate with a component depending on $t_{43}+t_{21}$, $t_{21}$, $t_{43}$, and $t_{32}$, in agreement with what we experimentally see in the time domain maps of the linear and the nonrephasing nonlinear signals. We do not find any terms oscillating at the rephasing frequency, or at $\Phi_{21}$. This could explain why the nonrephasing part is more strongly affected by the nonlinearity.\cite{Gregoire2017} However, this conclusion rests on a truncated expansion: by considering higher order terms, components oscillating at a combination of the phases could become important.\\
Empirically, it seems that our system can be intrinsically described by an expansion in power terms. The question remains if this is a correct assumption, and if so how we can determine the coefficients $\beta_i$ in Equation~\ref{eq:ExpansionSeries}. These coefficients hold the key to the prospect of disentangling the coherent from the incoherent contributions. A possible way to recover these coefficients could be to develop a circuit analysis, in analogy to what is routinely done for electrochemical impedance spectroscopy, for instance. By building an equivalent circuit of the system under investigation, it could be possible to extract the information needed to reconstruct Equation~\ref{eq:ExpansionSeries}. An important point that needs to be addressed is the universality of Equation~\ref{eq:ExpansionSeries} and, in a correlated but distinct way, of the coefficients $\beta_i$. Both the power terms in Equation~\ref{eq:ExpansionSeries} and the  coefficients could be directly related to the physical origin of nonlinear mixing, and in this case they could be different depending on the specific process, for example for an external electric response, or for bimolecular recombination versus Auger recombination. 
An indication that there are higher order terms playing a substantial role is given by the presence of higher order harmonics in the detected photocurrent signal. We use only two phase-locked pulses, and set the lock-in to pick up components oscillating at integer multiples of the reference frequency, $m\Omega_{ji}$, $m=1,2,3,4$. The results can be found in Fig.~\ref{fig:NHarmonics}, as a function of the fluence. The signal at the fundamental frequency is linear in a log-log scale for the range of fluences tested here. At low fluence the signal for higher order harmonics starts around two orders of magnitude below the signal at $\Omega_{ji}$, but the gap in the amplitudes closes at higher fluences. We can also notice a change in the slope of the amplitudes of $m\Omega_{ji}$, where above $0.1$\,$\mu$J/cm$^2$ the trend becomes sublinear. These higher order terms come either from a $mth$ coherent excited state or from higher order terms in the expansion of Equation \ref{eq:S_tot}. Although much lower than the fundamental signal, $m\Omega_{ji}$ signals are still relevant due to the extremely high signal-to-noise ratio typical of phase-sensitive techniques, and can easily mask the coherent signal or cause a misinterpretation of 2D data. It is therefore important to be able to detect the presence of incoherent mixing, by acquiring and carefully inspecting simultaneously both the nonlinear and linear maps in the time-domain.  

\section{Conclusions and outlook}
In this paper we used the electronic impedance of an external transimpedance amplifier circuit to simulate 
nonlinearity in the photocurrent signal  
from a model GaAs solar cell. We showed that in a real-case scenario the link between the phases of the coherent and incoherent contributions is not trivial, and we propose a way of recognizing the presence of incoherent mixing in 2D data by simultaneously acquiring the linear signals and analyzing their behaviour in the time domain. Incoherent mixing will cause the linear signal to oscillate also along the temporal axis associated to the pulse-pairs whose frequency the lock-in is looking at, and not only along the orthogonal one. We empirically derive a power series of the purely coherent response $\mathcal{S}_{coh}$ that generates terms oscillating at the nonrephasing frequency and at $\Phi_{43}$: these terms depend on $t_{43}+t_{21}$, $t_{21}$, $t_{43}$, and $t_{32}$, in agreement with the experimental evidence. Further investigation is needed to validate the reasonableness and universality of our hypothesis, and to unravel the connection between the proposed power series and the physical processes leading to nonlinear mixing. A possible way would be to develop an equivalent circuit modeling the system under study, that could take into account phenomenologically all the phenomena involved in the incoherent mixing. 

\section*{Supplementary Material}
Supplementary material include the details of the set up and of the GaAs solar cell. Also included are the maps in the time domain of the rephasing signal, and details on the simulation of the two pulse experiment for a GaAs solar cell. Lastly, SI include the calculations of $S^2_{co}$.

\begin{acknowledgments}
The authors are very grateful to Prof Ned Ekins-Daukes for providing the GaAs solar cells needed for this study. 
C.S.\ acknowledges support from the National Science Foundation (Grant DMR-1904293) and from the School of Chemistry and Biochemistry and the College of Science of Georgia Institute of Technology. 
\end{acknowledgments}


%

\newpage

\begin{center}
\Large
\textbf{Supporting Information}\\ \large
\textbf{Identifying incoherent mixing effects in the coherent two-dimensional photocurrent excitation spectra of semiconductors}\\
Bargigia~\etal
\end{center}
\normalsize

\section*{2D Spectroscopy Set-Up}
The laser source is a Pharos laser, Light Conversion. It emits 1030\,nm pulses at 300\,kHz, with an output power of 5\,W and pulse duration of $\sim220$\,fs. This laser pumps a non-collinear optical parametric amplifier (NOPA, Orpheus-N-2H from Light Conversion), operated at $1.65$\,eV (753\,nm). Pulses are compressed by a pulse shaper (Biophotonics Solution FemtoJock-P) in order to get transform-limited pulses of $\sim28$\,fs of Full-Width at Half Maximum at the sample position. Acousto-optic modulators impart to each of the four excitation pulses a modulation frequency of the order of $\sim200$\,MHz, while the frequency $\Omega_{21}$ and $\Omega_{43}$ are set at 4.5\,kHz and 13.4\, kHz, respectively. Phase-sensitive detection is achieved through a lock-in amplifier (Zurich Instruments HF2LI). A transimpedance amplifier (Zurich Instruments, HF2TA) is used to amplify and transduce the current generated in the solar cell to a voltage.    

\section*{G\lowercase{a}A\lowercase{s} Solar Cell}
\begin{figure}
    \centering
    \includegraphics[width=0.6\textwidth]{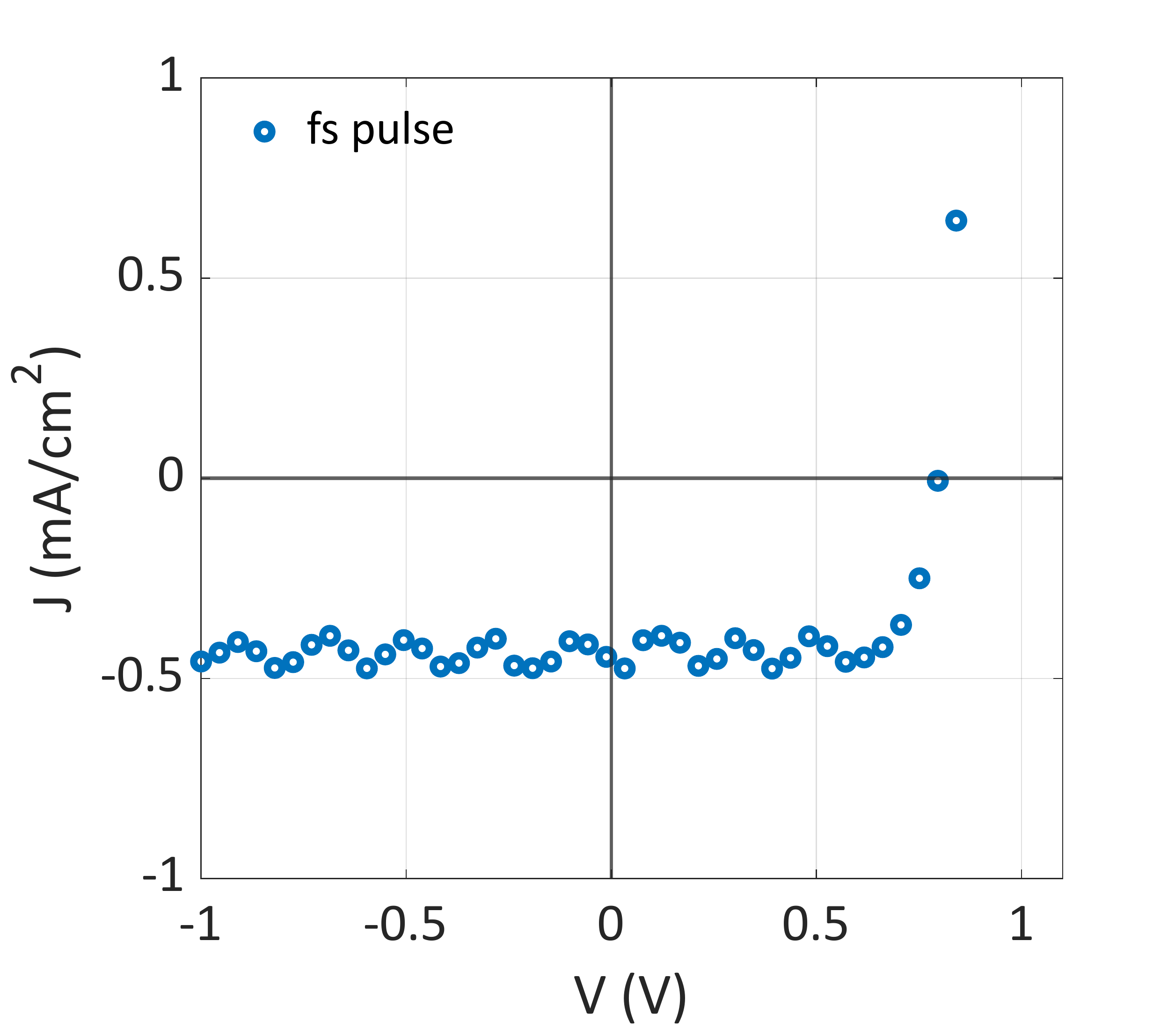}
    \caption{Current density of the solar cell under fs laser illumination. Fluence: $8 \times 10^{-2}$\,$\mu$J/cm$^2$.}
    \label{fig:JV}
\end{figure}

The solar cell was supplied by Professor Ned Ekins-Daukes. The device has a 0.4\,$\mu$m undoped GaAs layer. The short circuit current under 1-sun is approximately 220\,$\mu$A with open circuit voltage of $\sim$1\,V. The current density of the solar cell under fs laser illumination is reported in Fig.~\ref{fig:JV}. These data were acquired by connecting the solar cell to a Keithley source meter, under laser illumination. The fluence was $\sim 8 \times 10^{-2}$\,$\mu$J/cm$^2$. The open circuit voltage obtained under laser illumination is $\sim0.8$\,V, substantially lower than the nominal value as expected when nonlinear population mixing effects become dominant. The technical sheet of the solar cell is appended to this document. 

\section*{Voltage signal read by the lock-in}
\begin{figure*}
    \centering
    \includegraphics[width=17cm]{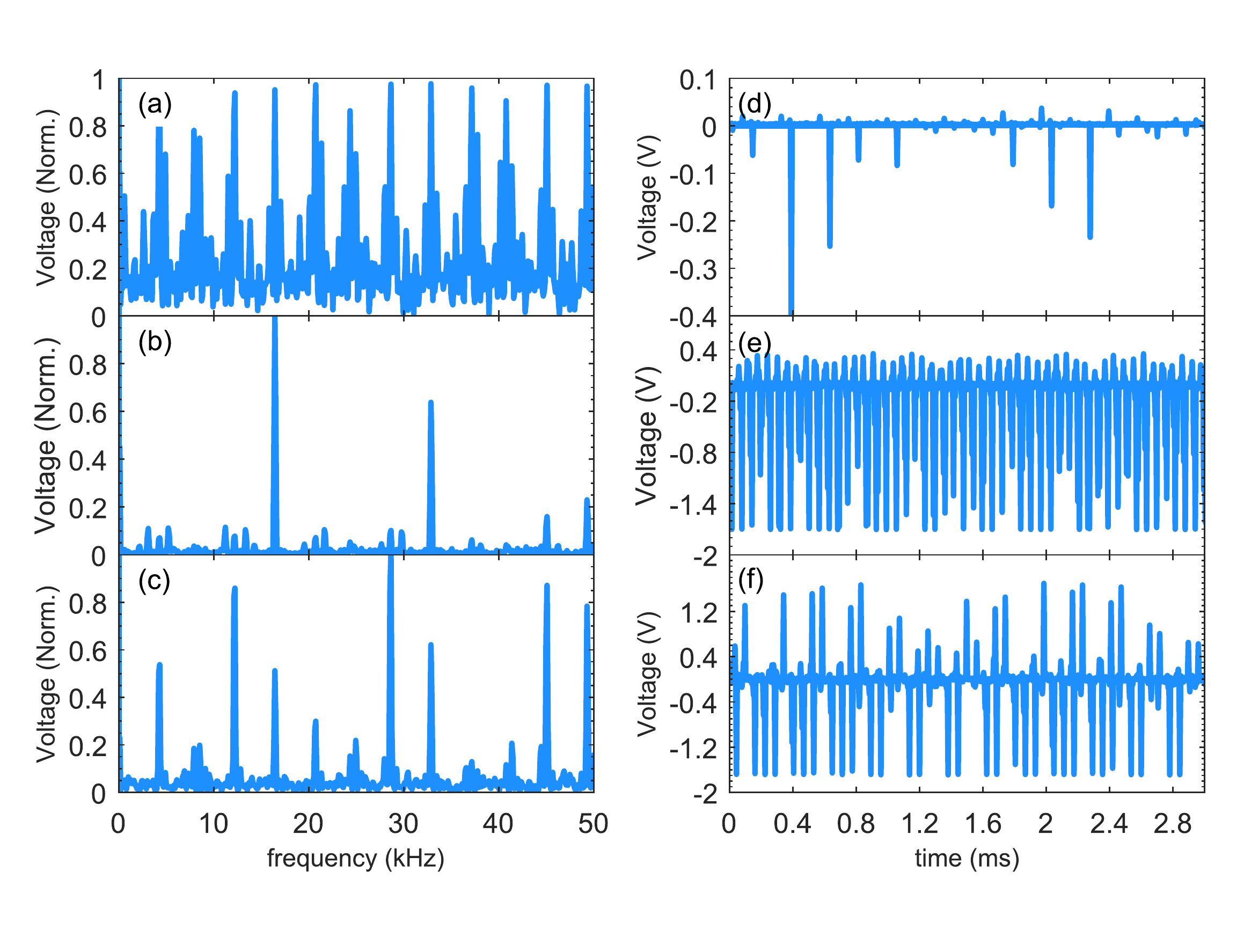}
    \caption{\textcolor{RoyalBlue}{On the left: f}requency spectra from the oscilloscope in-built in the lock-in amplifier for (a) $G_{TOT,1}$, (b) $G_{TOT,2}$, and (c) $G_{TOT,3}$; \textcolor{RoyalBlue}{On the right: t}ime domain data from the oscilloscope in-built in the lock-in amplifier for (d) $G_{TOT,1}$, (e) $G_{TOT,2}$, and (f) $G_{TOT,3}$. }
    \label{fig:LockIn}
\end{figure*}

Figure~\ref{fig:LockIn} shows the voltage signal generated at the output of the TIA as it is read by the lock-in amplifier, before any demodulation. The left column reports this voltage as a frequency spectrum (the fourier transform of the time domain data measured by the lock-in). As expected, for $G_{TOT,1}$ we see peaks at the AO frequencies 5.2\,kHz and 13.3\,kHz, at the difference frequency 8.1\,kHz, and at the sum frequency 18.5\,kHz. \textcolor{ForestGreen}{As we are using a pulsed excitation, with light pulses of $\sim28\,$fs (finite bandwidth) every $\sim 3.3\,\mu$s, the voltage signal output of the circuit is a train of voltage pulses with characteristic response time of the device imprinted on it. This will result in additional Fourier components in the  frequency domain, which are neither representative of the modulation frequencies nor the laser frequency. The weight of these additional frequencies is the result of the convolution between the response of the photovoltaic cell and the pulsed excitation, which is further subjected to a phase modulation. In addition, at low signal intensities, substantial contributions from the non-ideal contacts, signal oscillations from capacitive connectors manifest additional frequencies. }Despite these artefacts, it is clear that at $G_{TOT,2}$, new frequency components appear in the spectrum, and the distribution of the amplitudes changes. This effect is even more marked for $G_{TOT,3}$. The right column shows the data in the time domain.

\section*{Time-Domain Maps of the Rephasing Signal}
\begin{figure}
    \centering
    \includegraphics[width=0.6\textwidth]{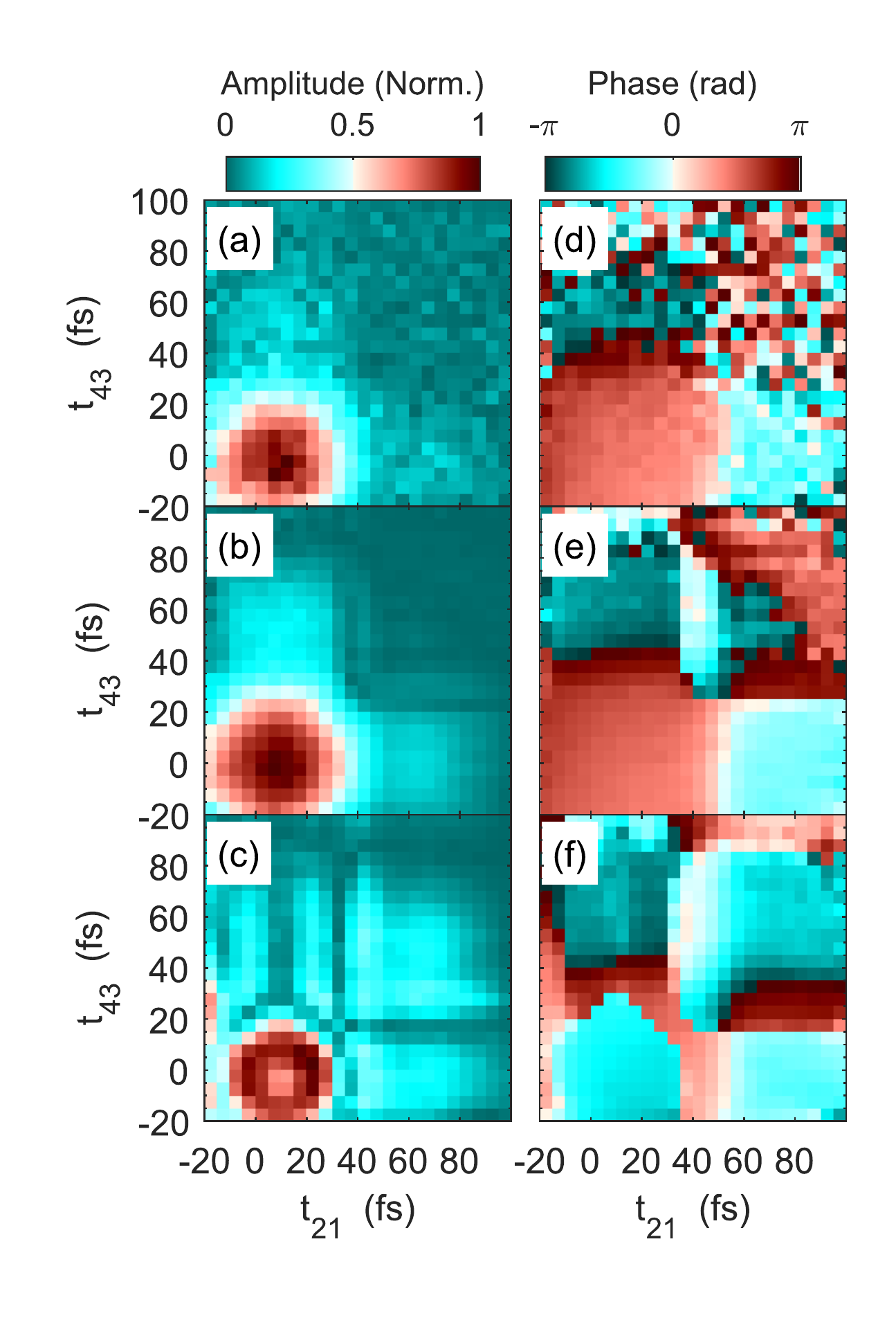}
    \caption{(a)--(c): Amplitude of the rephasing signal demodulated at $\Omega_{43} - \Omega_{21}$ for $G_{TOT,1}$, $G_{TOT,2}$, and $G_{TOT,3}$ respectively. (d)--(f): Phase of the rephasing signal demodulated at $\Omega_{43} - \Omega_{21}$ for $G_{TOT,1}$, $G_{TOT,2}$, and $G_{TOT,3}$.}
    \label{fig:reph_data_time}
\end{figure}

Maps in the temporal domain of the rephasing signal are reported in Fig.~\ref{fig:reph_data_time}.

\section*{Simulation of a two-pulse experiment}
The amplitude and phase of the photocurrent signal extracted from the GaAs solar cell were simulated usign Matlab with the following parameters. The instrumental response function $f(t_{ji})$ is assumed to be a Gaussian function with a FWHM of 28\,fs. The effective mass of the electron is $m^*=0.067\,m$, where \textit{m} is the mass of the electron. The spectral amplitude of the electric fields is considered identical for the two pulses, and it is assumed a Gaussian function centered at the central frequency of the pulses, $2 \pi c/(753 \, 10^{-9}\mbox{m})$\,rad/s. The dephasing rate is $\sim 0.008$\,fs$^{-1}$, obtained by fitting the anti-diagonal feature of the absolute nonrephasing signal at 1.63\,eV for the case of least nonlinearity, $G_{TOT,1}$. The direct bandgap of GaAs is usually set at 1.424\,eV. However, it is difficult to get this value experimentally from the absorption of this particular cell due to a possibly very extended Urbach tail. Furthermore, the design of the solar cell follows a layered structure, where doped layers sandwich undoped GaAs. We then expect a blueshift to occur, following a Berstein effect in the doped layers. The value chosen for the simulations, 1.6\,eV, is just an estimate and used to give a qualitative understanding of the expected signals.

\section*{Second Order Expansion of $\mathcal{S}_{coh}$}
As discussed in the paper, the total coherent response of the sample can be written as the superposition of the interferences between the wave packets created by pulse 1 and 2 ($n^{(2)}(t_{21})$), by pulse 3 and 4 ($n^{(2)}(t_{43})$), and the third-order wave packets $n^{(4)}(t_{ji},t_{lk};t_{kj})$. 
Let us write these contributions separately:
\begin{itemize}
    \item Interference signal $\braket{\psi_2|\psi_1}$:\\
    $S_{21}(t_{21},\Phi_{21}) = 2\Re\sum k_1 \, exp_{21} \, e^{i\Phi_{21}}$
    \item Interference signal $\braket{\psi_4|\psi_3}$:\\
    $S_{43}(t_{43},\Phi_{43}) = 2\Re\sum k_1 \, exp_{43} \, e^{i\Phi_{43}}$
    \item Interference signal $\braket{\psi_4|\psi_{321}}$:\\
    $S_{4}(t_{43},t_{21},\Phi_{SUM}) = 2\Re\sum_{a,b} k_2 \, exp_{a} \, e^{i\Phi_{SUM}}$
    \item Interference signal $\braket{\psi_{432}|\psi_{1}}$:\\
    $S_{1}(t_{43},t_{21},\Phi_{SUM}) = 2\Re\sum_{a,b} k_2 \, exp_{b} \, exp_{32} \, e^{i\Phi_{SUM}}$
    \item Interference signal $\braket{\psi_{421}|\psi_3}$:\\
    $S_{3}(t_{43},t_{21},\Phi_{DIFF}) = 2\Re\sum_{a,b} k_2 \, exp_{c} \, e^{i\Phi_{DIFF}}$
    \item Interference signal $\braket{\psi_{431}|\psi_2}$:\\
    $S_{2}(t_{43},t_{21},\Phi_{DIFF}) = 2\Re\sum_{a,b} k_2 \, exp_{c} \, exp_{32} \, e^{i\Phi_{DIFF}}$,
\end{itemize}
where we introduced the following terms to simplify the notation:
$k_1 = |\mu_{ag}|^2 \alpha^2(\omega_{ag})$, $exp_{21} = e^{-i\omega_{ag}t_{21}}$, $exp_{43}= e^{-i\omega_{ag}t_{43}}$, $k_2 = \mu_{ag}^2\mu_{bg}^2 \alpha^2(\omega_{ag})\alpha^2(\omega_{bg})$, $exp_a = e^{-i\omega_{ag}t_{43}-i\omega_{bg}t_{21}}$, $exp_b = e^{-i\omega_{ag}(t_{43}+t_{21})}$, $exp_{32} = e^{-i(\omega_{ag}-\omega_{bg}) \, t_{32}}$, $exp_c =  e^{-i\omega_{ag}t_{43}+i\omega_{bg}t_{21}}$. 
We considered that the spectral amplitudes of the pulses is exactly the same.
The coherent signal can then be written as:
\begin{equation}
    S_{coh} = S_{21} + S_{43} + S_4 + S_1 + S_3 + S_2.
\end{equation}
We can compute the second order of this signal by squaring the above terms:
\begin{equation}
    \begin{split}
    S_{21}^2(t_{21},\Phi_{21}) &= 4\Re\sum k_1^2 \, exp_{21}^2 \, e^{i2\Phi_{21}}, \\
    S_{43}^2(t_{43},\Phi_{43}) &= 4\Re\sum k_1^2 \, exp_{43}^2 \, e^{i2\Phi_{43}}, \\
    S_{4}^2(t_{43},t_{21},\Phi_{SUM}) &= 4\Re\sum_{a,b} k_2^2 \, exp_{a}^2 \, e^{i2\Phi_{SUM}},\\
    S_{1}^2(t_{43},t_{21},\Phi_{SUM}) &= 4\Re\sum_{a,b} k_2^2 \, exp_{b}^2 \, exp_{32}^2 \, e^{i2\Phi_{SUM}},
        \end{split}
\end{equation}
\begin{equation}
    \begin{split}
    S_{3}^2(t_{43},t_{21},\Phi_{DIFF}) &= 4\Re \sum_{a,b} k_2^2 \, exp_c^2 e^{i2\Phi_{DIFF}},\\
    S_2^2(t_{43},t_{21},\Phi_{DIFF}) &= 4\Re\sum_{a,b}k_2^2 \, exp_c^2 \, exp_{32}^2 \, e^{i2\Phi_{DIFF}},\\
    2 \, S_{21}(t_{21},\Phi_{21})S_{43}(t_{43},\Phi_{43}) &= 8\Re\sum k_1^2 \, exp_{21} \, exp_{43} \, e^{i\Phi_{SUM}},\\
    2\,S_{21}(t_{21},\Phi_{21})\,S_{4}(t_{43},t_{21},\Phi_{SUM}) &= 8\Re \sum_{a,b} k_1\,k_2\,exp_{21}\,exp_a\,e^{i(\Phi_{43}+2\Phi_{21})},\\
    2\,S_{21}(t_{21},\Phi_{21})\,S_{1}(t_{43},t_{21},\Phi_{SUM}) &= 8\Re \sum_{a,b} k_1\,k_2\,exp_{21}\,exp_b\,exp_{32}\,e^{i(\Phi_{43}+2\Phi_{21})},\\
    2\,S_{21}(t_{21},\Phi_{21})\,S_{3}(t_{43},t_{21},\Phi_{DIFF}) &= 8\Re \sum_{a,b} k_1\,k_2\,exp_{21}\,exp_c\,e^{i\Phi_{43}},\\
    2\,S_{21}(t_{21},\Phi_{21})\,S_{2}(t_{43},t_{21},\Phi_{SUM}) &= 8\Re \sum_{a,b} k_1\,k_2\,exp_{21}\,exp_c\,exp_{32}\,e^{i\Phi_{43}},\\
    2\,S_{43}(t_{43},\Phi_{43})\,S_{4}(t_{43},t_{21},\Phi_{SUM}) &= 8\Re \sum_{a,b} k_1\,k_2\,exp_{43}\,exp_a\,e^{i(2\Phi_{43}+\Phi_{21})},\\
    2\,S_{43}(t_{43},\Phi_{43})\,S_{1}(t_{43},t_{21},\Phi_{SUM}) &= 8\Re \sum_{a,b} k_1\,k_2\,exp_b\,exp_{43}\,exp_{32}\,e^{i(2\Phi_{43}+\Phi_{21})},\\
    2\,S_{43}(t_{43},\Phi_{43})\,S_{3}(t_{43},t_{21},\Phi_{SUM}) &= 8\Re \sum_{a,b} k_1\,k_2\,exp_{43}\,exp_c\,e^{i(2\Phi_{43}-\Phi_{21})},\\
    2\,S_{43}(t_{43},\Phi_{43})\,S_{2}(t_{43},t_{21},\Phi_{DIFF}) &= 8\Re \sum_{a,b} k_1\,k_2\,exp_{43}\,exp_c\,exp_{32}\,e^{i(2\Phi_{43}-\Phi_{21})},\\
    \end{split}
\end{equation}
\begin{equation*}
    \begin{split}
        2\,S_{4}(t_{43},t_{21},\Phi_{SUM})\,S_{1}(t_{43},t_{21},\Phi_{SUM}) &= 8\Re \sum_{a,b} k_2^2\,exp_b\,exp_{a}\,exp_{32}\,e^{i(2\Phi_{SUM})},\\
    2\,S_{4}(t_{43},t_{21},\Phi_{SUM})\,S_{3}(t_{43},t_{21},\Phi_{DIFF}) &= 8\Re \sum_{a,b} k_2^2\,exp_a\,exp_{c}\,e^{i(2\Phi_{43})},\\
    2\,S_{4}(t_{43},t_{21},\Phi_{SUM})\,S_{2}(t_{43},t_{21},\Phi_{DIFF}) &= 8\Re \sum_{a,b} k_2^2\,exp_c\,exp_{a}\,exp_{32}\,e^{i(2\Phi_{43})},\\
    2\,S_{1}(t_{43},t_{21},\Phi_{SUM})\,S_{3}(t_{43},t_{21},\Phi_{DIFF}) &= 8\Re \sum_{a,b} k_2^2\,exp_b\,exp_{32}\,e^{i(2\Phi_{43})},\\
    2\,S_{1}(t_{43},t_{21},\Phi_{SUM})\,S_{2}(t_{43},t_{21},\Phi_{DIFF}) &= 8\Re \sum_{a,b} k_2^2\,exp_c\,exp_{b}\,exp^2_{32}\,e^{i(2\Phi_{43})},\\
    2\,S_{3}(t_{43},t_{21},\Phi_{DIFF})\,S_{2}(t_{43},t_{21},\Phi_{DIFF}) &= 8\Re \sum_{a,b} k_2^2\,exp_c^2\,exp_{32}\,e^{i(2\Phi_{DIFF})}.
    \end{split}
\end{equation*}

\newpage
\begin{figure*}[htpb]
    \centering
    \includegraphics[width=\textwidth]{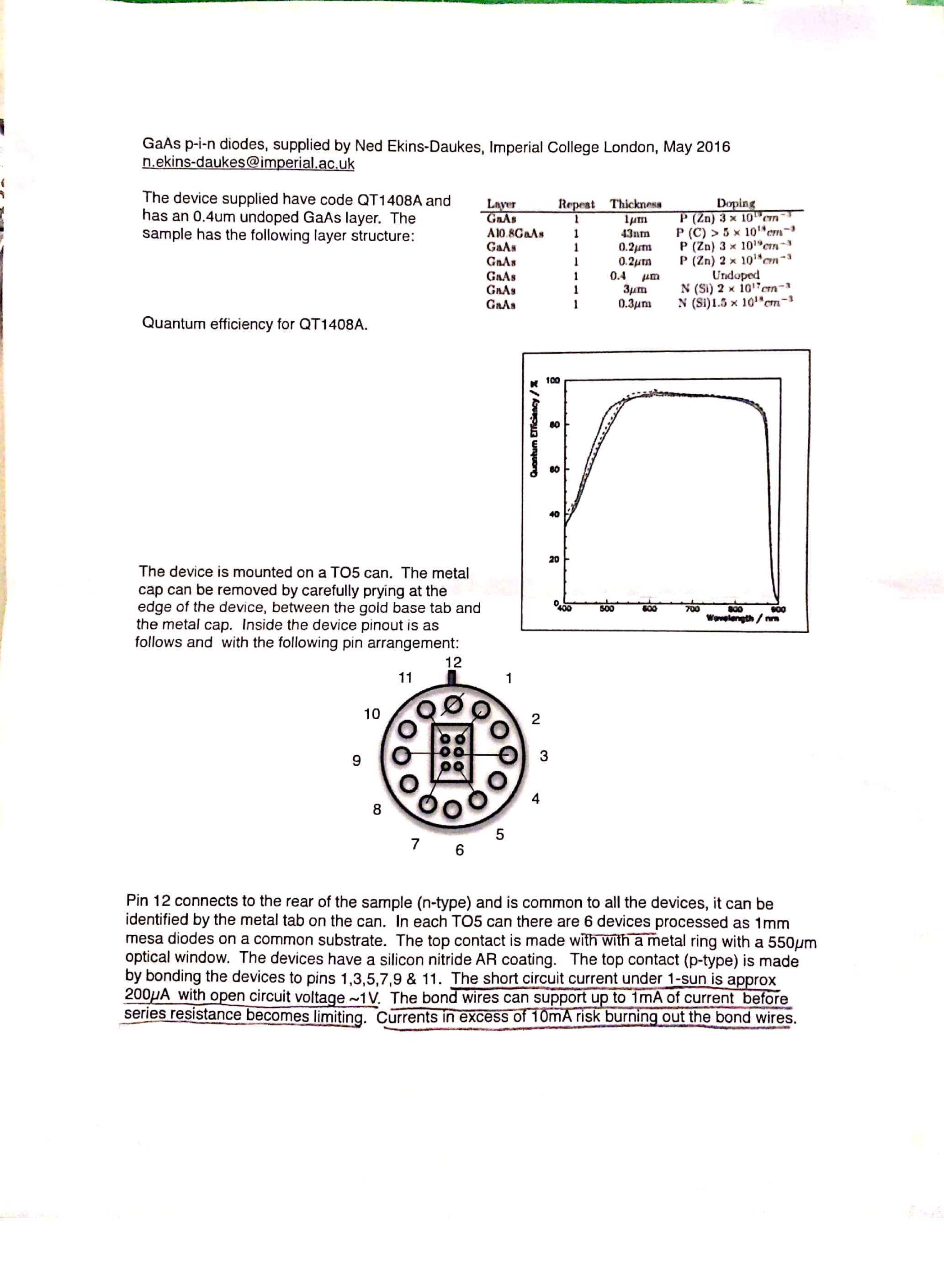}
    \label{fig:tech_sheet}
\end{figure*}

\end{document}